\documentclass{aa}
\usepackage[varg]{txfonts}
\bibpunct{(}{)}{;}{a}{}{,} 
\usepackage{hyperref} 
\usepackage{breqn}
\hypersetup{colorlinks=true,citecolor=blue}

\begin{document}

\title{On the influence of equilibrium tides on transit-timing variations of close-in super-Earths}

\subtitle{I. Application to single-planet systems and the case of K2-265 b}

\author{G. O. Gomes\inst{1,2}
  \and E. Bolmont\inst{1}
  \and S. Blanco-Cuaresma\inst{3}}

\institute{Observatoire de Gen\`eve, Universit\'e de Gen\`eve, Chemin Pegasi 51, CH-1290 Sauverny, Switzerland \\ \email{Gabriel.De-Oliveira@etu.unige.ch}
  \and Instituto de Astronomia, Geof\'isica e Ci\^encias Atmosf\'ericas, IAG-USP, Rua do Mat\~ao 1226, 05508-900 S\~ao Paulo, Brazil
  \and Harvard-Smithsonian Center for Astrophysics, 60 Garden Street, Cambridge, MA 02138, USA}

\date{Received 22 December 2020 / Accepted ???}

\abstract{With the current growth in the discovery of close-in low-mass exoplanets, recent works have been published with the aim to discuss the influences of planetary interior structure parameters on both the shape of transit light curves as well as variations on the timing of transit events of these planets. One of the main aspects explored in these works is the possibility that the precession of the argument of periapsis caused by planetary tidal interactions may lead to unique effects on the transit light curves of the exoplanets, such as the so-called transit-timing variations (TTVs).

In this work, we investigate the influence of planetary tidal interactions on the transit-timing variations of short-period low-mass rocky exoplanets. For such purpose, we employ the recently-developed creep tide theory to compute tidally-induced TTVs. We implement the creep tide in the recently-developed Posidonius N-body code, thus allowing for a high-precision evolution of the coupled spin-orbit dynamics of planetary systems. As a working example for the analyses of tidally-induced TTVs, we apply our version of the code to the K2-265 b planet. We analyse the dependence of tidally-induced TTVs with the planetary rotation rate, uniform viscosity coefficient and eccentricity. Our results show that the tidally-induced TTVs are more significant in the case where the planet is trapped in non-synchronous spin-orbit resonances, in particular the 3/2 and 2/1 spin-orbit resonant states. An analysis of the TTVs induced separately by apsidal precession and tidally-induced orbital decay has allowed for the conclusion that the latter effect is much more efficient at causing high-amplitude TTVs than the former effect by 2 - 3 orders of magnitude. We compare our findings for the tidally-induced TTVs obtained with Posidonius with analytical formulations for the transit timings used in previous works, and verified that the results for the TTVs coming from Posidonius are in excellent agreement with the analytical formulations. These results show that the new version of Posidonius containing the creep tide theory implementation can be used to study more complex cases in the future. For instance, the code can be used to study multiplanetary systems, in which case planet-planet gravitational perturbations must be taken into account additionally to tidal interactions to obtain the TTVs.}

\maketitle 

\section{Introduction}

The measurement of transit-timing variations (TTVs) has been shown to be an invaluable tool in the field of exoplanetary detection and characterization \citep{Agol2005}. In multi-planet systems, for instance, the gravitational interaction between planets leads to deviations from strictly Keplerian orbits. These deviations in turn lead to a shift in the timing of planetary transit events. The TTVs effects are usually investigated by directly running N-body dynamical simulations. The advantage of such an approach lies in the fact that the results are valid for arbitrary eccentricities and inclinations of the planets. Several studies have employed the technique of TTVs analyses to improve the mass measurement of disturbing (non-transiting) planetary companions, as well as their orbital parameters \citep{Nesvorny2008,Lithwick2012,Nesvorny2016,Agol2020}.

Although planet-planet interactions are among the main effects leading to high-amplitude TTVs, any interaction causing a drift of a planet from its unperturbed Keplerian orbit can induce TTVs. For very close-in planets, for instance, general relativity-induced apsidal precession and tidally-induced orbital decay (see e.g. discussions in \citealt{Ragozzine2009}) may be strong enough to cause significant variations of the orbital elements of the planets and lead to observable TTVs. The detection of such TTVs caused by these disturbing forces depends on the magnitude of the induced TTVs compared to the precision of the instruments measuring transit events.

In the field of study of tidal interactions and subsequent spin-orbit evolution, the parameter which is responsible for dictating the response to tidal stress is the complex Love number, which is the ratio of the additional gravitational potential induced by internal mass redistribution of the planet due to tidal deformation to the external tidal potential created by the perturber \citep{Love1911}. Thus, provided that the tidally-induced TTVs are big enough to be detected, matching the observed TTVs with the values estimated by numerical simulations of the tidal evolution of the system would allow us to estimate the complex Love number of the bodies, thus providing us with invaluable information regarding e.g., the interior structure of the planets and their rotational state.

In a recent work, \citet{Patra2017} analysed transit and occultation data of WASP-12 b (see e.g., \citealt{Hebb2009,Haswell2018} and references therein) and showed that the planet could be undergoing an orbital decay process corresponding to a mean period decay of $\dot{P} = -29 \pm 3 \ \textrm{ms yr}^{-1}$, which would be compatible with a stellar quality factor of the order $2 \times 10^5$. However, it was pointed out by the authors that apsidal precession induced by planetary tidal deformation could provide the same effects on the transit curves when compared to the orbital decay process induced by stellar tides, in which case the Love number of the planet would be of the order $k_p = 0.44 \pm 0.10$. The confusion between the two effects could only be solved if occultation data were analysed additionally to transit data, in which case the contribution from apsidal precession gives an opposite sign to the occultation timing variation curves when compared to orbital decay. In a more recent work, \citet{Yee2020} have analysed new transit and occultation measurements for the same system and verified that a model for the transit timings considering orbital decay is more likely to explain the TTV curve of the planet when compared to a model considering apsidal precession.

Although the recent discussions raised by e.g., \citet{Ragozzine2009}, \citet{Patra2017} and \citet{Yee2020} have paved the way for future investigations regarding the tidally-induced TTVs, few studies have been dedicated to an exploration of these effects for low-mass close-in rocky exoplanets. The reason for the lack of studies for such class of exoplanets is linked to the fact that only a few of them had been discovered before the results of photometric surveys coming from missions such as Kepler \citep{Youdin2011,Fressin2013} and TESS \citep{Sullivan2015}. However, with the recent boom in the discovery and characterization of close-in rocky super-Earths and the improvement in the precision of the transit timing measurements, an investigation of the potential effects of planetary tidal interactions on TTVs is necessary. This may be a powerful approach at assessing the planets interior structure (see e.g., \citealt{Bolmont2020a}) as well as other parameters, such as their current rotational state (where the rotational state of the planet indeed depends on its interior structure parameters, such as its viscosity and density). 

In the field of study of tidal interactions on rocky super-Earths, several works have aimed at a description of the equilibrium tide and its subsequent orbital evolution scenarios by employing advanced rheological models to describe the response of these exoplanets to tidal stress (see e.g., \citealt{Efroimsky2012, FM2013,Tobie2019,Walterova2020} and references therein). However, to the best of our knowledge, none of these advanced rheological models have been consistently implemented in high-precision and open-source N-body codes\footnote{We emphasize that a self-consistent formulation for the coupled spin-orbit and shape evolution considering a Maxwell rheology for the planet was presented in \citet{Correia2014}, which was extended to study (i) three-body exoplanetary systems by e.g., \citet{Rodriguez2016} and \citet{Delisle2017} and (ii) N-body systems by \citet{Correia2019}.}, which would allow for a multi-planet simulation of tidally-induced TTVs to a good degree of accuracy. In a recent work, \citet{Bolmont2020a} have investigated the magnitude of tidally-induced TTVs for the TRAPPIST-1 multi-planet system by employing the implementation of a classical approach of tidal interactions based on the Constant Time Lag (CTL) formulation \citep{Mignard1979}. However, this specific approach does not allow for a full exploration of the TTVs considering all the nuances of equilibrium tides interactions in rocky bodies, such as the temporary entrapment in spin-orbit resonances for moderately high values of eccentricities and planetary viscosity (see e.g., \citealt{Makarov2013} and references therein). Such aspects, which are consistently reproduced using more complex rheological models, may indeed play an important role on the orbital evolution of the planets and, consequently, lead to significant changes in the tidally-induced TTVs when compared to scenarios obtained by employing classical theories of equilibrium tides, such as the CTL approach.

Taking into account the discussions presented in e.g. \citet{Delisle2017} and \citet{Bolmont2020a} related to the potentially observable effects of tides on TTVs as well as the lack of an investigation of such topic using a more sophisticated tidal interaction model for the planet, we revisit, in this work, the topic of tidally-induced TTV analyses in the frame of the creep tide theory \citep{FM2013,Folonier2018}. For this purpose, we implement the tidal interactions model in the recently-developed N-body code called Posidonius\footnote{The new version of the code including the implementation of the creep tide theory is available for download in https://github.com/gabogomes/posidonius/tree/posidonius-creep, and the original Posidonius website is https://www.blancocuaresma.com/s/posidonius} \citep{Blanco2017,Bolmont2020a}.

The creep tide theory \textrm{describes the} response of bodies to tidal stress by supposing a Newtonian creep equation for the extended body's shape evolution. This equation is the result of a linear approximation of the Navier-Stokes equation for a low-Reynolds-number flow (i.e., for cases where viscous forces are much more significant than inertial forces, see \citealt{FM2013}). We consider the formulation of the theory for homogeneous bodies whose rotation axes are perpendicular to the orbital plane (i.e., the zero-obliquity case). The equations ruling the spin and orbit evolution of the planets are shown to be easy to implement in the Posidonius code, and the response of bodies to tidal stress depends on only one parameter, which is the uniform viscosity coefficient of the body. 

As a real example of a tidally-induced TTV analysis, we study the K2-265 b super-Earth \citep{Lam2018}. We explore the influence of tidally-induced TTVs as a function of several parameters, such as the planet's viscosity, eccentricity and rotational state. We also present a broad analysis of the amplitudes of tidally-induced TTVs as a function of other parameters, such as the planet's mass and radius values and its semi-major axis. Such broad analysis allows us to identify which are the main parameters leading to bigger influences on the TTV values. Finally, we also provide some examples of recently-discovered systems for which tidally-induced TTVs would have the highest amplitudes for a given number of transit events. Such systems are the candidates for which tidally-induced TTVs can be more easily detected by the instruments measuring transit events, such as the TESS telescope. 

This work is organized as follows: In Section\,\ref{sec:sec2}, we present the main results of the creep tide theory which were used for the implementation in the Posidonius code and briefly give some details regarding the numerical setup used in the calculations of the TTVs. In Section\,\ref{sec:sec3}, we investigate the magnitudes of tidally-induced TTVs for the K2-265 b planet considering the influences of tuning the parameters presenting the most significant uncertainties in their determined values (i.e., the eccentricity, rotational state and viscosity of the planet). We also compare the results for the TTVs obtained by using Posidonius with the ones obtained by employing some analytical approximations to generate transit timing curves. In Section\,\ref{sec:sec4}, we perform a broad exploration of tidally-induced TTVs as a function of both orbital and physical parameters of the star and the planet. In Section\,\ref{sec:sec5} we present the discussions of the work, and Section\,\ref{sec:concl} summarizes our conclusions.

\section{Numerical setup}
\label{sec:sec2}

In this section, we provide the essential equations of the creep tide theory which were used in the implementation in the Posidonius code. We also present some additional information regarding the numerical integrator which provides the better precision for the calculations of transit timing events. 

\subsection{The creep tide theory}

\label{sec:creep}

\begin{figure}[!h]
\centering
\includegraphics[height=140pt,width=250pt]{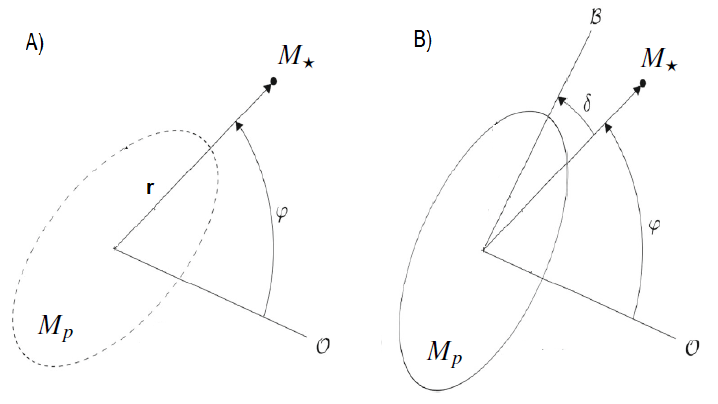}
\caption{Scheme of the geometrical setup considered in the framework of the creep tide theory, partially reproduced from \citet{Folonier2018}. On the left, we have the pure static case in which the primary always points towards the companion, which is at a distance $\vec{r}$ from the primary. On the right, we have the case in which the response of the primary is delayed due to the primary's non-zero viscosity value. In both panels, $\mathcal{O}$ is the reference direction used to measure the true anomaly ($\varphi$).}
\label{fig:fig0}
\end{figure}

The creep tide theory considers the deformations of an extended body (hereafter the primary) due to the disturbing gravitational potential due to a point-mass companion. The time evolution of the primary's figure is described by three parameters: the equatorial deformation of the body ($\mathcal{E}_{\rho}$), polar oblateness ($\mathcal{E}_z$) and the angle giving the orientation of the tidal bulge ($\delta$), as it is described in the scheme in Fig.\,\ref{fig:fig0}. These three parameters are evolved by employing a Newtonian creep equation, which is an approximate solution of the Navier-Stokes equation in the case of a low-Reynolds-number flow (see \citealt{FM2013,Folonier2018} for more details).

Employing the expression for the potential of the resulting triaxial figure of the primary, we compute the forces and the torque acting on the companion. While the radial and ortho-radial components of the force (namely $\vec{F}_r$ and $\vec{F}_o$) are used to compute the orbital evolution of the system, the reaction to the torque acting on the companion is used to compute the rotational evolution of the primary. The calculations are straightforward and they entail \citep{Folonier2018}

\begin{equation}
\vec{F}_{\textrm{r}} = \left[ -\frac{9}{10} \frac{GM_pM_{\star}R_p^2}{r^4} \mathcal{E}_{\rho} \cos 2 \delta - \frac{3}{5} \frac{GM_pM_{\star}R_p^2}{r^4} \mathcal{E}_z \right] \vec{r} ,
\label{eq:f-rad}
\end{equation}
\begin{equation}
\vec{F}_{\textrm{o}} = \left[ \frac{3}{5} \frac{GM_pM_{\star}R_p^2}{r^4} \mathcal{E}_{\rho} \sin 2 \delta \right] (\vec{z} \times \vec{r}) ,
\label{eq:f-ortho-rad}
\end{equation}
\begin{equation}
\dot{\Omega} = -\frac{ 3 G M_{\star} }{2 r^3} \mathcal{E}_{\rho} \sin 2 \delta .
\label{eq:omegadot-complete}
\end{equation}

The unit vectors $\vec{r}$ and $\vec{z} \times \vec{r}$ point towards the radial and ortho-radial directions, respectively, with $\vec{z}$ being a unit vector pointing to the same direction of the orbital angular momentum vector. The parameters $G$, $M_p$, $R_p$, $M_{\star}$ are the gravitational constant, mass and radius of the planet and mass of the star, respectively. 


To compute the time evolution of $\mathcal{E}_{\rho}$, $\mathcal{E}_{z}$ and $\delta$, we employed the analytical solutions for these parameters in the frame of the constant rotation rate approximation \citep{Gomes2019}, which read 

\begin{equation}
\mathcal{E}_{\rho} \cos 2 \delta = \bar{\epsilon}_{\rho} \sum _{k \in \mathbb{Z}} \mathcal{C}_k [\gamma \cos \phi _k - (2 \Omega - k n) \sin \phi _k] ,
\label{eq:bx_analytic}
\end{equation}
\begin{equation}
\mathcal{E}_{\rho} \sin 2 \delta = \bar{\epsilon}_{\rho} \sum _{k \in \mathbb{Z}} \mathcal{C}_k [\gamma \sin \phi _k + (2 \Omega - k n) \cos \phi _k] ,
\label{eq:by_analytic}
\end{equation}
\begin{equation}
\mathcal{E}_{z} = \bar{\epsilon}_z + \frac{\bar{\epsilon}_{\rho}}{2} \left[ \sum_{k \in \mathbb{Z}} \mathcal{Z}_k (k^2n^2\sin k \ell + \gamma ^2 \cos k \ell) \right] ,
\label{eq:bz_analytic}
\end{equation}
where
\begin{equation}
\phi _k = k \ell - 2 \varphi ,
\label{eq:p1}
\end{equation}
\begin{equation}
\mathcal{C}_k = \frac{\gamma E_{2,2-k}}{\gamma ^2 + (2 \Omega - k n)^2} ,
\label{ref:eqc1}
\end{equation}
\begin{equation}
\mathcal{Z}_k = \frac{E_{0,k}}{k^2n^2 + \gamma ^2} ,
\label{ref:eqc2}
\end{equation}
\begin{equation}
\bar{\epsilon}_{z} = \frac{5\Omega^2R_p^3}{4GM_p} ,
\end{equation}
\begin{equation}
\bar{\epsilon}_{\rho} = \frac{15}{4} \frac{M_{\star}}{M_p} \left(\frac{R_p}{a}\right)^3 ,
\label{eq:eqprol}
\end{equation}
\begin{equation}
\gamma = \frac{R_pgd}{2\eta} .
\label{eq:visc-gamma}
\end{equation}

A list containing the meaning of each symbol in Eqs.\,\ref{eq:p1}-\ref{eq:visc-gamma} is presented in Table\,\ref{table0}.

The Cayley coefficients appearing in Eqs.\,\ref{ref:eqc1} and \ref{ref:eqc2} (namely, $E_{2,2-k}$ and $E_{0,k}$) are eccentricity-dependent\footnote{We comment that the expansion in eccentricity coefficients converges only for eccentricities up to $\approx 0.67$. For a formulation allowing arbitrary eccentricity values, see e.g., \citet{Correia2014,Folonier2018}.}. Their values can be computed numerically, by means of the integral
\begin{equation}
E_{q,k} (e) = \frac{1}{2\pi \sqrt{1-e^2}} \int_{0}^{2\pi} \frac{a}{r} \cos [q\varphi + (k-q) \ell] d\varphi .
\label{eq:cayley}
\end{equation}

	\begin{table}
  \begin{center}
    
    \begin{tabular}{ccccc} 
      \hline
     Symbol & Meaning \\ 
      \hline
      
      $\ell$ & mean anomaly \\
      $\varphi$ & true anomaly \\
      $n$ & orbital mean-motion \\
      $\gamma$ & planetary relaxation factor \\
      $\eta$ & planetary uniform viscosity \\
      $d$ & planetary mean density \\
      $g$ & planetary surface gravity \\
      $E_{q,k}$ & Cayley coefficient \\
      \hline
      
    \end{tabular}
    \caption{Meaning of the parameters presented in Eqs.\,\ref{eq:p1}-\ref{eq:visc-gamma}.} 
    \label{table0}
  \end{center}
\end{table}

Although the evaluation of high-order Cayley coefficients is relatively straightforward by employing Eq.\,\ref{eq:cayley}, we used polynomial expressions for these coefficients to order $e^{7}$ (see \citealt{FM2015}, Appendix B.4, Tables 1 and 2) in all the applications performed in this work, since such an approximation has already been shown to be sufficient to calculate the tidal evolution for eccentricity values up to approximately 0.4 (see \citealt{FM2013}, Appendix).

We emphasize that the constant rotation rate approximation presented in this section can be used in our study since the librations of the rotation rate which ensue in spin-orbit resonant states do not significantly change the orbital evolution of the system (and, consequently, the TTVs). However, they may have important contributions to the overall energy dissipated in the body (see discussions in \citealt{Folonier2018}). Additionally, we point out that the constant rotation rate solution is not a secular solution of the creep tide equations, and short-period oscillations of the shape coefficients are present. The shape evolution of the primary depends implicitly on the mean and true anomalies, see Eqs.\,\ref{eq:bx_analytic}-\ref{eq:bz_analytic}.

\subsection{Implementation in Posidonius}

To explore the tidally-induced TTVs with a high-degree of numerical precision, we implemented the creep tide theory following the scheme presented in Sec.\,\ref{sec:creep} (see Eqs.\,\ref{eq:f-rad}-\,\ref{eq:omegadot-complete}) in the recently-developed Posidonius code \citep{Blanco2017}. 

We compared the performance and numerical precision of two integrators: WHFast \citep{Rein2015b}, which is a symplectic integrator considering a fixed time-step scheme for the numerical integration of the equations of motion, and IAS15 \citep{Rein2015a}, which considers an adaptive time-step control scheme based on Gauss-Radau spacings for the numerical integration of the equations of motion. We performed some numerical experiments and verified that, although WHFast is approximately 20 times faster than the IAS15 integrator, the latter provides a bigger numerical precision for very short timescales of evolution, which is essential at evaluating the TTVs properly. Thus, we considered the IAS15 in all simulations regarding the TTV calculations, where we set $max(\Delta t) = 0.0005$ days, with $max(\Delta t)$ being the maximum value of the allowed time-step of the numerical integration scheme. Such value of $max(\Delta t)$ provides a precision of approximately $10^{-4} \ \textrm{s}$ for the TTVs, where such precision value was estimated by analysing the TTVs of a single-planet case with no disturbing forces other than the point-mass gravitational interaction between the star and the planet. More information regarding the numerical validation and performance of the code can be found in the Appendix.

\subsection{Calculation of TTVs}

To calculate the TTVs of each transit event, we employ the same procedure described in \citet{Bolmont2020a}. The procedure is briefly described below, for the sake of clarity.

We firstly perform a numerical simulation employing the Posidonius code, including the effects of the tidal interactions. Then, we use the data coming from the time right before and right after the point in which the planet crosses a reference direction (in our case, we chose the $X$ axis to be the reference direction) and linearly interpolate the orbit to find the exact time for which $X=0$. Then, we perform a least-square linear fit using the $\tens{LinearRegression}$ algorithm from the $\tens{linear_}  \tens{model}$ package of $\tens{scikit}$-$\tens{learn}$ (see \citealt{Pedregosa2011}). This process allows us to write the transit mid-time associated with each transit event as $T = a n_T + b$, where $T$ and $n_T$ are the transit mid-times and transit number, respectively, and the coefficients $a$ and $b$ are the best-fit orbital period and transit mid-time of the first transit event, respectively. Afterwards, we compute the differences between the transit times coming from the linear fit (i.e., the calculated transit times, $C_i$) and the transit times resulting from the numerical simulations employing Posidonius (i.e., the observed transit times, $O_i$). The difference between these two quantities (namely, $O_i - C_i$) are the TTVs.

We comment that the above procedure to obtain the tidally-induced TTVs is slightly different from the one employed by \citet{Bolmont2020a}. In our procedure, we did not calculate the TTV difference between the pure N-body case and the case considering the tidal interactions. The absence of such part of the procedure is due to the fact that, since we are dealing with a single-planet system, the expected TTV for the pure N-body case is zero (i.e., there are no TTVs for a non-perturbed Keplerian orbit).

\section{Application to K2-265 b}
\label{sec:sec3}

In this section, we analyse the tidally-induced TTV for K2-265 b. We analyse the effect of tuning some parameters of the planet presenting the most significant uncertainties (namely, its viscosity, eccentricity and rotation rate) on the amplitudes of the TTVs. We also forecast tidally-induced TTVs considering bigger numbers of transit data than the currently available amount of reported transit events.

\subsection{Currently available data for the system}
\label{sec:k2265b}

The K2-265 b super-Earth was discovered by \citet{Lam2018}. The authors used photometry data from the K2 mission and high-precision radial velocity measurements from the High Accuracy Radial velocity Planet Searcher (HARPS, \citealt{Mayor2003}), and they estimated the planet radius and mass with an uncertainty of $6\%$ and $13\%$, respectively. The planet orbits a G8V bright star with an apparent magnitude of $V=11.1$. The estimated radius and mass of the planet are $R = 1.71 \pm 0.11 \ R_{\oplus}$ and $6.54 \pm 0.84 \ M_{\oplus}$, which corresponds to a mean density of $7.1 \pm 1.8 \ \textrm{g cm}^{-3}$. Such mean density value is typical of rocky Earth-like planets. 

Regarding the orbital parameters of the planet, \citet{Lam2018} obtained $a = 0.03376 \pm 0.00021 \ \textrm{AU}$ and $e = 0.084 \pm 0.079$. An estimation of the age of the system (henceforth referred to as $\tau$) was also made available by \citet{Lam2018} by combining observational data and stellar evolution tracks coming from the Dartmouth stellar evolution code. The resulting estimated age for the system is $\tau = 9.7 \pm 3.0 \ \textrm{Gyr}$.

The complete set of parameters which were used to perform our numerical experiments is shown in Table\,\ref{table1}.

	\begin{table}[!h]
  \begin{center}
    
    \begin{tabular}{ccccc} 
      \hline
     Parameter & Value \\ 
      \hline
      
      Planet mass $(M_{\oplus})$ & $6.54 \pm 0.84$ \\
      Planet radius $(R_{\oplus})$ & $1.71 \pm 0.11$\\
      Planet mean density $(\textrm{g cm}^{-3})$ & $7.1 \pm 1.8$\\
      Semimajor axis $(\textrm{AU})$ & $0.03376 \pm 0.00021$ \\
      Orbital period (days) & $2.369172 \pm 0.000089$ \\
      Eccentricity & $0.084 \pm 0.079$ \\
      Stellar mass $(M_{\star})$ & $0.915 \pm 0.017$ \\
      Stellar rotation period (days) & $32 \pm 10$ \\
      Age $(\textrm{Gyr})$ & $9.7 \pm 3.0$ \\
      
      \hline
      
    \end{tabular}
    \caption{Physical and orbital parameters for K2-265 b and its host star after \citet{Lam2018}. The values represent the median of the posterior and $68.3\%$ credible interval. For more details on the methods of determination of each parameter, see \citet{Lam2018}.} 
    \label{table1}
  \end{center}
\end{table}

\subsection{Estimation of the uniform viscosity value}

\label{sec:sec31}

To study the tidally induced TTV for K2-265 b, firstly we have to constrain the range of values for the uniform viscosity coefficient which are consistent with both the putative rocky composition of the planet as well as the current eccentricity estimations available. For such purpose, we used a secular code to calculate tidal evolution scenarios (see Appendix) to derive a mathematical relation linking the uniform viscosity coefficient to the timescale of orbital circularization of the planet, namely $\tau _{\textrm{circ}}$. Differently from other works where $\tau _{\textrm{circ}}$ is defined as the inverse of the coefficient multiplying the expression for $de/dt$ (see e.g., \citealt{Ballard2014} and references therein), we define $\tau _{\textrm{circ}}$ as the time it would take for a planet with an initial eccentricity of $e_0 = 0.4$ to reach a current eccentricity which is smaller than $10^{-3}$. We emphasize that the choice of the value of $e_0=0.4$ is arbitrary, and that we have tested other scenarios considering different values of $e_0$ and verified that such parameter does not play an important role on the results for $\tau _{\textrm{circ}}$, since the orbital circularization process becomes slower for the small-eccentricity regime.

\begin{figure}
\centering
\includegraphics[height=180pt,width=230pt]{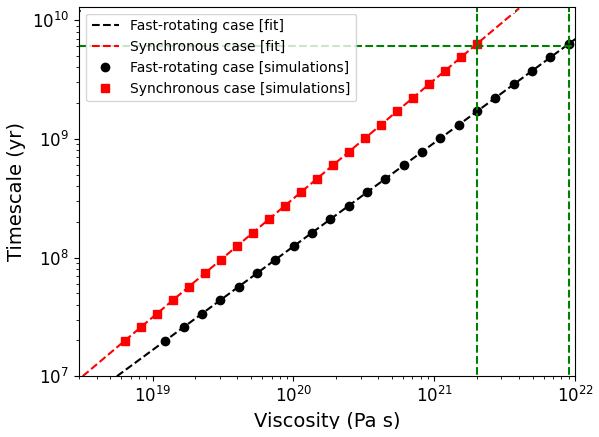}
\caption{Timescales of orbital circularization for K2-265 b. The red squares (resp. black dots) are the data obtained from the numerical integrations employing the secular code, for the initially synchronous (resp. fast-rotating) planet case. The red (resp. black) dashed curves are the corresponding least-squares fittings of the results coming from the secular code experiments. The green dashed vertical line on the left (resp. right) is the lower value of the viscosity for which we can reproduce the planet's current range of eccentricity values for the initially synchronous (resp. fast-rotating) planet case, with $\tau  = \tau _{\textrm{circ}}=6.7 \ \textrm{Gyr}$.}
\label{fig:fig1}
\end{figure}

Fig.\,\ref{fig:fig1} shows the timescale-viscosity relations resulting from our tidal evolution numerical experiments. The black (resp. red) curve shows the results for the initially fast-rotating (resp. synchronous) planet scenario. The corresponding mathematical relations between $\eta$ and $\tau _{\textrm{circ}}$ are

\begin{equation}
    \eta _{\textrm{FR}} = (5.3 \times 10^{10}) \tau _{\textrm{circ}} ^{1.15} ,
    \label{eq:fr}
\end{equation}
\begin{equation}
    \eta _{\textrm{S}} = (3.2 \times 10^{11}) \tau _{\textrm{circ}}  ,
    \label{eq:sync}
\end{equation}
where $\eta _{\textrm{FR}}$ (resp. $\eta _{\textrm{S}}$) is the viscosity for the initially fast-rotating (resp. synchronous) planet case (see black and red curves in Fig.\,\ref{fig:fig1}).

From the results shown in Fig.\,\ref{fig:fig1}, we estimate that a viscosity of the order $10^{22} \ \textrm{Pa s}$ or higher is needed for the current eccentricity of the planet to be within the limits estimated by \citet{Lam2018} (i.e., for $0.005 \leq e \leq 0.15$), supposing that the system's age is of the order $6 - 12 \ \textrm{Gyr}$ (see green dashed lines in Fig.\,\ref{fig:fig1}). This viscosity value is in agreement with the recent estimations of \citet{Bolmont2020b} for homogeneous rocky Earth-like planets.

By propagating the eccentricity and age uncertainties to the viscosity estimation presented here, we conclude that the minimum value of the planet's uniform viscosity is $\eta _{\textrm{min}} \approx 2 \times 10^{21} \ \textrm{Pa s}$ (where $\eta _{\textrm{min}}$ is the viscosity value for which the planet would have a current eccentricity of $0.005$ considering an age of $6.7 \ \textrm{Gyr}$). Although an estimation of the maximum viscosity value of the planet is not possible since any scenario with $\eta > 10^{22} \ \textrm{Pa s}$ could lead to a current eccentricity between 0.005 and 0.15, we comment that recent estimations of the viscosity in planetary interiors do not exceed approximately $10^{24} \ \textrm{Pa s}$, and such threshold value is only attained for very high-pressure regimes in the interior of low-mass planets (see \citealt{Tobie2019}, Table 2). 

Finally, we mention that the differences in the exponents of $\tau_{\textrm{circ}}$ (see Eqs.\,\ref{eq:fr} and\,\ref{eq:sync}) is a consequence of the fact that the tidal response of the body depends on the forcing frequency of the tidal interaction, the latter being defined as $\chi _{lmpq} = \left\vert (l-2p+q)n - m \Omega \right\vert$ (see e.g., \citealt{Efroimsky2012,Tobie2019}). Since in the initially synchronous scenario the forcing frequency is (approximately) constant throughout the entire orbital evolution process, we expect a linear relationship between the timescale and the viscosity. However, in the initially fast-rotating case, the rotational evolution of the planet leads to the excitation of different terms of the tidal forcing frequencies which have different amplitudes, thus leading to a non-linear relationship between the timescale of orbital circularization and the viscosity of the planet.

\subsection{The TTVs}

In this section, we investigate the TTVs for K2-265 b, where we analyse the influence of tidal interactions as well as other potentially significant effects such as general relativity (henceforth referred to as GR) and stellar rotation. To that end, we employed the Posidonius code, where the creep tide was implemented following the equations presented in the previous section. For the sake of clarity, we separate the analyses in several subsections to discuss the influence of tuning different parameters individually. With the exception of the subsection aimed at discussing the influence of tuning the uniform viscosity coefficient, we considered $\eta = 10^{22} \ \textrm{Pa s}$ in all other simulations. Such value is consistent with both our orbital circularization timescales and recent estimations of the viscosity in planetary interiors (e.g silicate mantles with pressure values of $P > 25 \ \textrm{GPa}$ and shear to bulk moduli ratio of $\mu / K$ between 0.63 and 0.90 P/K, see \citealt{Tobie2019,Bolmont2020b}). 

\subsubsection{Interplay between apsidal precession and orbital decay}

The first aspect to be discussed regarding the TTVs is the relative contribution of orbital decay and apsidal precession. Firstly, we recall the expressions for the transit timings supposing two cases: (i) a circular orbit undergoing orbital decay at rate $da/dt$, and (ii) an eccentric orbit undergoing apsidal precession at rate $d\omega / dt$. The expression for the transit timing in the first case reads (see e.g., \citealt{Ragozzine2009})

\begin{equation}
    t_{\textrm{tra}} (N) = t_0 + NP + \frac{N^2}{2} \left[ \frac{6\pi^2a^2}{G(M+m)}\right] \left( \frac{da}{dt} \right),
    \label{transit:dadt}
\end{equation}
where $P$ is the mean orbital period of the planet.

For the second case, we have, to third-order expansion in the eccentricity (see \citealt{Gimenez1995,Ragozzine2009} for higher-order expansions)
\begin{dmath}
    t_{\textrm{tra}} (N) = t_0 + NP_S + \frac{P_a}{\pi} \left[ e \left( \cos \omega _N - \cos \omega _0 \right) + \frac{3}{8} e^2 \left( \sin 2\omega _N - \sin 2\omega _0 \right) + \frac{1}{6} e^3 \left( \cos 3\omega _N - \cos 3\omega _0 \right) \right] ,
    \label{transit-dw}
\end{dmath}
where
\begin{equation}
\omega _N = \omega _0 + \frac{d\omega}{dN} N,
\end{equation}
with $t_0$, $\omega _0$ being the time and argument of periapsis value of the first reported transit, respectively, and $N$ corresponding to the transit number. $P_S$ and $P_a$ correspond to the sidereal and anomalistic periods, respectively (see e.g., \citealt{Patra2017} for more details).

As it can be seen from Eqs.\,\ref{transit:dadt} and \ref{transit-dw}, the contributions coming from apsidal precession and orbital decay to the TTVs have the same dependence on the transit number, i.e., both contributions give an $N^2$ dependence on the TTV curves (to first order in $\cos \omega _N$).

To disentangle between the effects of apsidal precession and orbital decay on the TTVs, we compared the TTVs obtained by employing two procedures: (i) the data coming from the numerical experiments using the Posidonius code, and (ii) synthetic transit curves, generated by employing Eqs.\,\ref{transit:dadt} and \ref{transit-dw}. For the synthetic transit curves using Eqs.\,\ref{transit:dadt} and \ref{transit-dw}, we used the values of $\dot{\omega}$ and $\dot{a}$ ensuing from the Posidonius numerical experiment. In the case of a K2-265 b with $e=0.15$ and $\Omega = n$, for instance, we have $\dot{\omega} = 0.046 \ \textrm{deg / yr}$ and $\dot{a} = -1.91 \times 10^{-11} \ \textrm{AU / yr}$. We mention that, to investigate the contribution of the apsidal precession to the TTVs, we included the effects of general relativity and stellar rotation in addition to tidal interaction. We verified that the most significant contribution to $\dot {\omega}$ comes from general relativity, which accounts for approximately $93\%$ of the total value of $\dot{\omega}$. Stellar rotation accounts for approximately $6\%$ of the value of $\dot{\omega}$ (considering the smallest possible rotation period for the star according to \citealt{Lam2018} and  a stellar $k_2$ of 0.1) and planetary tides account for approximately $1\%$. Stellar tides play the least important role with less than $1\%$ of contribution to $\dot{\omega}$.

\begin{figure}
\centering
\includegraphics[height=120pt,width=255pt]{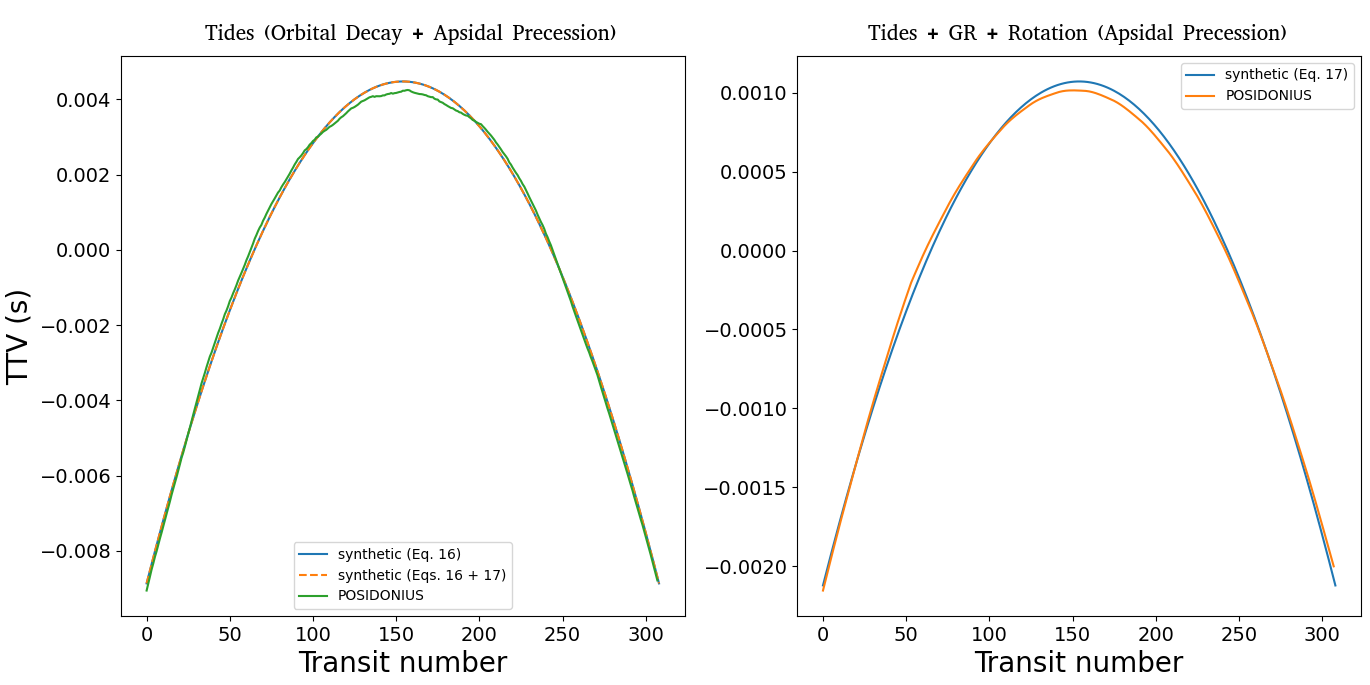}
\caption{TTVs for K2-265 b, considering $\eta = 10^{22} \ \textrm{Pa s}$, $e=0.15$ and a synchronous rotation rate for the planet. The panel on the left was obtained considering only tidal interactions, while the case on the right was obtained by considering tides, rotation and general relativity (the contribution coming from tidal decay to the TTVs on the yellow curve were filtered out by directly subtracting the TTVs generated by employing Eq.\,\ref{transit:dadt}). The blue curves are synthetic TTV curves and the yellow curves correspond to the results coming directly from Posidonius.}
\label{fig:fig9}
\end{figure}

Fig.\,\ref{fig:fig9} shows the results coming from two experiments: one simulation carried out considering only tidal interactions (panel on the left), and one simulation considering general relativity, rotation and tidal interactions (panel on the right). The contribution of tidally-induced orbital decay to the TTVs were filtered out of the yellow curve on the right panel of Fig.\,\ref{fig:fig9} by directly removing the TTVs generated using Eq.\,\ref{transit:dadt}. The total integration time was 2 years (corresponding to approximately 307 transit events for K2-265 b). For the case on the left, we verified that the synthetic curve generated using Eq.\,\ref{transit:dadt} (which corresponds to the blue curve) gives a TTV amplitude of the order $0.012 \ \textrm{s}$, while the synthetic curve generated using only Eq.\,\ref{transit-dw} (i.e., considering only the contribution of tidally-induced apsidal precession to the TTVs) gives a TTV amplitude of approximately $0.0001 \ \textrm{s}$. The negligible role of tidally-induced apsidal precession to the TTVs can be verified by comparing the blue solid curve and the yellow dashed curve on the left panel. It can be seen that the addition of tidally-induced apsidal precession effects on the TTVs does not change significantly the TTV curve, and the yellow and blue curves on the left panel are almost identical. However, general relativity plays a non-negligible role on the TTV, with GR-induced apsidal precession being able to produce a TTV amplitude of approximately $20\%$ of the amplitude obtained by considering the tidally-induced orbital decay (compare green curve on the left panel and yellow curve on the right panel in Fig.\,\ref{fig:fig9}).

Although general relativity has been shown to be non-negligible in the context of the TTVs, its relevance was analysed for the case of a synchronous planet with $\eta = 10^{22} \ \textrm{Pa s}$ and $e=0.15$. It is highly unlikely for a rocky planet with such viscosity and eccentricity values to be in a synchronous rotation rate scenario. Since tidally-induced orbital decay is bigger for non-synchronous spin-orbit resonant states, we expect that GR becomes negligible when considering other spin-orbit resonances for the planet. We will now analyse how these spin-orbit resonances affect the TTVs.

\subsubsection{The role of eccentricity and planet's rotation}

As it was discussed in several recent works regarding tidal interactions in rocky bodies, planets with a relatively high viscosity may undergo entrapment in spin-orbit resonances provided that the eccentricity of the planet is bigger than a threshold value and its past rotation was such that $\Omega \gg n$ (e.g., see \citealt{Correia2014,FM2015} and references therein). Considering the case of the K2-265 b exoplanet with a uniform viscosity of $\eta = 10^{22} \ \textrm{Pa s}$, several spin-orbit resonant states are stable if we consider the possible eccentricity values for the planet (namely, $0.005 \leq e \leq 0.15$). 

\begin{figure}
\centering
\includegraphics[height=210pt,width=255pt]{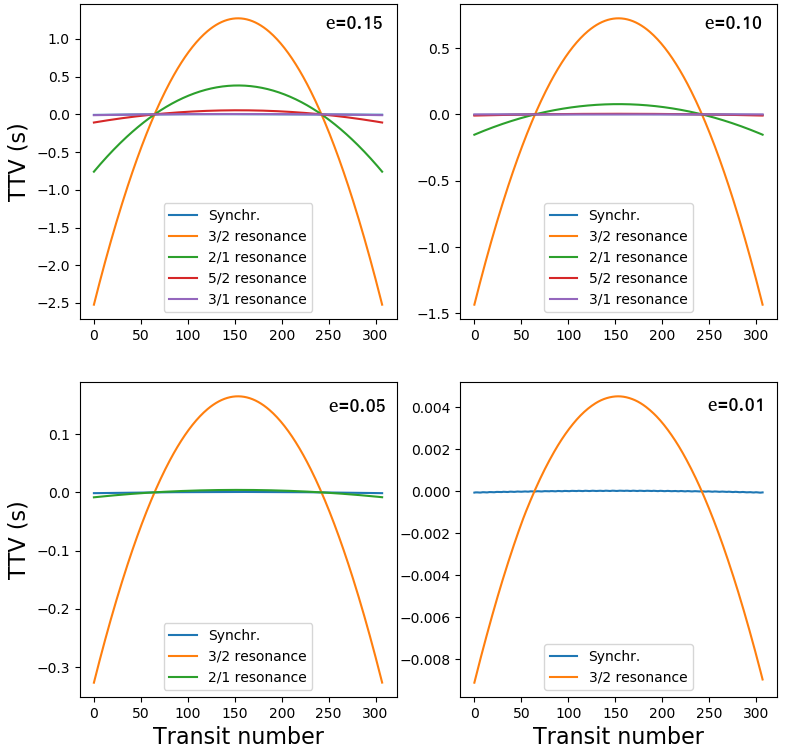}
\caption{Tidally induced TTV for a timespan of 2 years (corresponding to 308 transit events), considering several spin-orbit resonances and four values of eccentricity, as indicated in the top right corner of each panel.}
\label{fig:fig2}
\end{figure}

Due to the existence of several stable resonant scenarios, we calculated tidally-induced TTVs for the planet for four values of eccentricity (which are within the limit values established by \citealt{Lam2018}) and the corresponding spin-orbit stable configurations for each value of eccentricity, where we mention that increasing the eccentricity value leads to the onset of possible higher-order stable spin-orbit resonant states (to see details regarding the determination of stable spin-orbit resonances, see e.g., \citealt{Correia2014,FM2015,Gomes2019} and references therein). The results of our analysis are shown in Fig.\,\ref{fig:fig2}. In the figure, we can deduce both the influence of the eccentricity as well as the rotation rate of the planet on the amplitude of the TTVs (henceforth referred to as $\textrm{amp(TTV)}$), with 
\begin{equation}
\textrm{amp(TTV)} = \textrm{max(TTV)} - \textrm{min(TTV)} . 
\end{equation}

By comparing the yellow curves in each panel of Fig.\,\ref{fig:fig2}, it can be seen that the eccentricity value strongly affects the tidally-induced TTVs. We have verified that $\textrm{amp(TTV)}$ scales with $e^{\alpha}$, where $\alpha = 1, 2$ and $4$ for the synchronous, 3/2 and 2/1 spin-orbit resonances, respectively. Moreover, $\alpha$ increases monotonically with $\Omega / n$ (the latter being usually referred to as the order of the spin-orbit resonance). At this point, it is worth mentioning that the resulting linear dependence of the amplitude of the TTVs with the eccentricity, in the synchronous rotating planet case, is consistent with the predictions of \citet{Ragozzine2009}. We did not perform a study regarding the eccentricity dependence of the amplitudes of the TTVs for higher-order spin-orbit resonances (such as the 5/2 and 3/1 resonances) since such spin-orbit resonances are only maintained at either high eccentricity values with $e > 0.15$ (which are incompatible with the eccentricity estimations of K2-265 b) or very big viscosity values (which are inconsistent with our estimations of Sec.\,\ref{sec:sec31}). 

Another aspect which can be seen from Fig.\,\ref{fig:fig2}, by comparing, for instance, the curves on the top left panel, is that tidally induced TTVs are negligible for the synchronous rotation and high-order spin-orbit resonant states (e.g., the 5/2 and 3/1 resonances) when compared to the 3/2 and 2/1 resonances (see yellow and green curves on each panel).

\subsubsection{Impact of the uniform viscosity coefficient}

Although the combination of eccentricity and age estimations of the system has allowed for a reasonable estimation of the planet's viscosity, we must analyse the influence of tuning the viscosity value of the planet on the tidally induced TTVs, since the viscosity of the planet was estimated with a  given uncertainty. For such purpose, we consider viscosity values within the uncertainties discussed in the Section\,\ref{sec:sec31}.

\begin{figure}[!h]
\centering
\includegraphics[height=190pt,width=230pt]{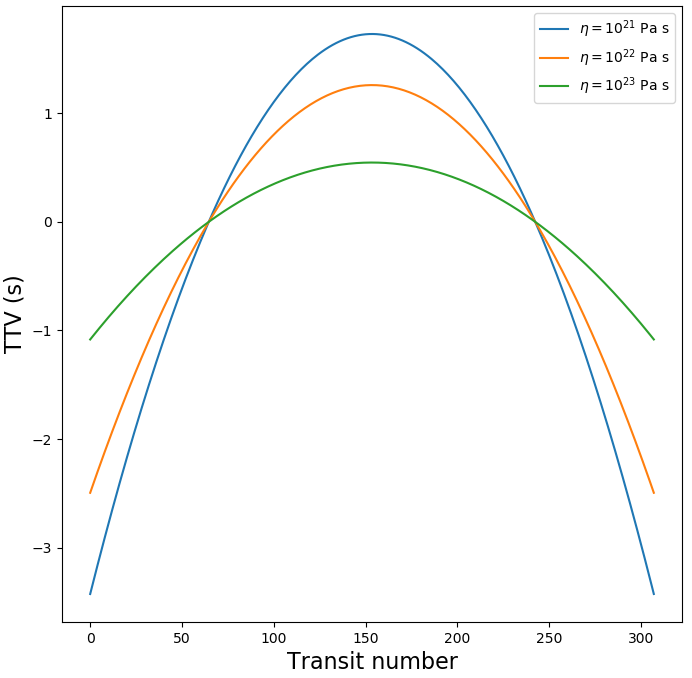}
\caption{Effect of the viscosity value on the tidally induced TTV for K2-265 b with $e=0.15$, $\Omega / n = 1.5$ and a total transit number of 2 years. We considered variations of $\eta$ between the minimum value (coming from eccentricity evolution timescale estimations), and the maximum value (coming from the reference values of \citealt{Tobie2019}, which were obtained by solving the internal structure equations for the mantle of planets composed by high-pressure silicates).}
\label{fig:fig4}
\end{figure}

Fig.\,\ref{fig:fig4} shows the TTVs curves for three numerical experiments where we tuned the viscosity value of the planet. We can see that the viscosity is the parameter playing the least important role when compared to the influence of the eccentricity and rotation rate on the amplitudes of the tidally-induced TTVs. In Fig.\,\ref{fig:fig4}, it can be seen that by increasing the viscosity by a factor of $100$, the amplitude of the tidally-induced TTVs decreases by a factor of approximately $3$. The influence of the viscosity value on the TTV amplitude is even weaker for higher-order spin-orbit resonances, as it was verified by some numerical experiments for the 2/1 and 5/2 spin-orbit resonances.

\subsection{Conditions for potentially observable TTVs}

\begin{figure}[!h]
\centering
\includegraphics[height=210pt,width=250pt]{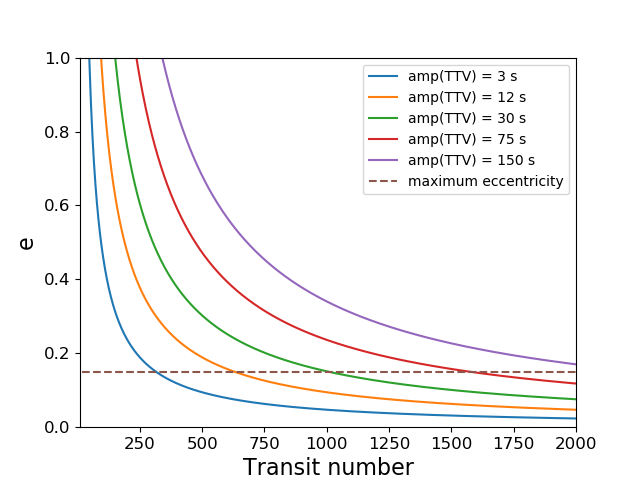}
\caption{Relation between eccentricity and transit number giving the amplitudes of TTVs indicated in the labels. In all cases, we considered the 3/2 spin-orbit resonance, which gives the highest amplitudes of the TTVs when compared to all the other rotational configurations. The maximum transit number shown in the figure (i.e., 2000 transits) corresponds to approximately 13 years of transit events. More discussions are presented in the main text.}
\label{fig:fig5}
\end{figure}

Since we have analysed the dependency of the tidally-induced TTVs with the parameters of the K2-265 system carrying the biggest uncertainties, we can now establish for which conditions we would have the biggest amplitude of tidally-induced TTVs (i.e., the scenarios leading to the most easily detectable tidally-induced TTVs with the smallest amount of transit data). Fig.\,\ref{fig:fig5} shows the results concerning such analysis for the specific case of the 3/2 spin-orbit resonant state, which is the resonant state presenting the biggest TTVs. The curves in the graph give the eccentricity and transit number values leading to the corresponding amplitudes of the TTVs (see labels on the upper right corner in the figure). The dashed brown line corresponds to the maximum value of the eccentricity of K2-265 b estimated by \citet{Lam2018}.

From the results shown in Fig.\,\ref{fig:fig5}, we can conclude that approximately 6 years (corresponding to approximately 1000 transit events) of transit data would be necessary for a TTV amplitude of the order 30 s to be obtained, considering that the planet's eccentricity value is the maximum possible value corresponding to the estimations of \citet{Lam2018}. However, if a precision of approximately 10 s for transit measurements is achieved (which is the case for the TRAPPIST-1 system, see discussions in \citealt{Bolmont2020a} and \citealt{Agol2020}), we predict that approximately 3.5 years of transit data would suffice to identify tidally-induced TTVs. The follow-up of transit photometry data in the future would then be an essential tool to experimentally identity such effects.

It is worth mentioning that we performed some numerical experiments to evaluate TTVs induced by the stellar tide as well as the stellar flattening resulting from the stellar rotation (where the stellar rotation period of 32.2 days was used in our simulations, following \citealt{Lam2018}). It was verified that these effects are much smaller than tidally-induced TTVs by the planet even in the case of a homogeneous star (i.e., with a fluid Love number of $k_f = 3/2$). An analysis of the influence of the general relativity on the induced TTVs was also performed. We also verified that these effects are negligible when compared to the planetary tidally-induced TTVs.

Although the results presented in this section show that the follow-up of transit data for the K2-265 b could potentially lead to observable TTVs in the near future, we will now perform a broad exploration of the tidally-induced TTVs as a function of the system's orbital and physical parameters to have a more comprehensive view of which are the most important parameters influencing the TTVs.

\section{Broad exploration of the parameters space}

\label{sec:sec4}

In the previous section, we have analysed tidally-induced TTVs for the K2-265 system and concluded that, with the current amount of available data, it is not yet possible to detect tidally-induced TTVs. In this section, we will explore the influence of other parameters on the values of the tidally-induced TTVs with the goal to determine for which exoplanetary systems we would be able to detect such effects with a smaller amount of transit data.

\begin{figure}[!h]
\centering
\includegraphics[height=220pt,width=250pt]{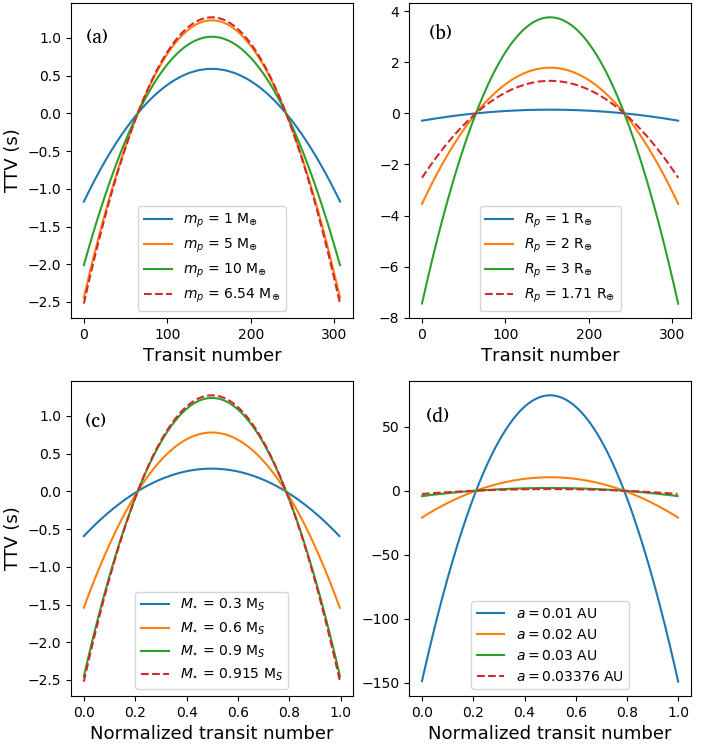}
     \caption{Broad exploration of the TTV as a function of four parameters: (a) planetary mass, (b) planetary radius, (c) stellar mass and (d) semi-major axis values. We considered a total timespan of 2 years in all the simulations. In each panel of the figure, we tuned individually each parameter and considered the nominal values of the K2-265 b system for the remaining ones (see Table\,\ref{table1}). Moreover, in all cases shown here, we considered a 3/2 spin-orbit resonant state for the planet as well as an eccentricity of 0.15, in order to maximize the effects of the TTVs.}
     \label{fig:fig6}
\end{figure}

Fig.\,\ref{fig:fig6} shows the results regarding the exploration of the tidally-induced TTVs as a function of four parameters (semi-major axis, planet mass, planet radius and stellar mass, see labels and caption of Fig.\,\ref{fig:fig6}). In all panels, the dashed red line corresponds to the case of the K2-265 b planet using the parameters of Table\,\ref{table1}. For the bottom panels, we used the normalized transit number to perform the plots, since changing the stellar mass and semi-major axis leads to considerable differences in the orbital period, which in turn causes alterations in the total transit number, even though the total time of evolution remains the same (namely, 2 years). We can clearly see that the most important parameters influencing the amplitudes of the TTVs are the stellar mass and the semi-major axis (see panels c and d in the figure). The planetary radius value also strongly affects the amplitude of the TTVs. However, since super-Earths are believed to have radii between 1 and 1.8 Earth radii according to recent discussions (see e.g., \citealt{Fulton2017}), the experiments performed in panel b for $R_p = 2 \ R_{\oplus}$ and $R_p = 3 \ R_{\oplus}$ (see yellow and green curves) are merely exploratory. 

The factor playing the least important role on the TTVs is the planetary mass. Tuning such parameter by a factor 10 leads to a difference in the TTVs of a factor 3. Moreover, we emphasize that changing the planetary mass leads to a direct change in the relaxation factor value, since the planetary mass directly affects the planet mean density value (see Eq.\,\ref{eq:visc-gamma}) and its mean equatorial prolateness. Thus, the influence of tuning the planetary mass on the tidally-induced TTVs is much more complex than the influences of the other three factors which were considered in this section, and a proportionality relation between the amplitude of the TTVs and the tuning of the planetary mass may not be possible (see yellow and green curves in the panel a of Fig.\,\ref{fig:fig6}).

\section{Discussions}

\label{sec:sec5}

In this section, we discuss several aspects of the results obtained in Sections \ref{sec:sec3} and \ref{sec:sec4}. For the sake of clarity, the discussions are separated in subsections.


\subsection{TTVs for K2-265 b}

Regarding our numerical experiments of the TTVs for K2-265 b, we verified that the most important factors ruling the amplitudes of the TTVs are the eccentricity and spin-orbit resonance. Regarding the spin-orbit resonance influence upon TTVs, we have verified that although synchronous rotation leads to small-amplitude TTVs, planets in low-order spin-orbit resonances (especially the 3/2 and 2/1 resonances) may present relatively high amplitudes of tidally-induced TTVs, where the main component causing the tidally-induced TTVs for non-synchronous rotation cases is the orbital decay of the planet. Other effects such as tidally-induced apsidal precession and GR-induced apsidal precession were verified to produce TTVs two to three orders of magnitude smaller than the TTVs caused by tidally-induced orbital decay. The uniform viscosity coefficient was shown to cause a relatively small variation on the amplitudes of TTVs when compared to the effects of tuning the rotation and eccentricity values.

In what concerns the predictions for future TTVs considering more transit data, we have shown that tidally-induced TTVs may be able to cause deviations of transit timings of the order 30 - 70 s considering a 10-year timespan. These results indicate that tides can be an important source of additional TTVs to be included when modelling transit data using a longer baseline in the case of close-in planetary systems.

\subsection{Potential confusion with other effects generating TTVs}

As it was discussed in Section\,\ref{sec:sec3}, the TTVs induced by tidal interactions are much more significant when the planet is trapped in a non-synchronous spin-orbit resonance, where the main component being responsible for the TTVs is the orbital decay of the planet. However, when the planet is trapped in a synchronous rotation, the TTV induced by orbital decay is less significant, and the TTV induced by the apsidal precession has a non-negligible role on the total TTV. The most significant effect causing apsidal precession-induced TTVs is the GR, with $\dot{\omega}^{\textrm{(GR)}}$ being at least one order of magnitude bigger than any other effects causing apsidal precession.

\begin{figure}[!h]
\centering
\includegraphics[height=120pt,width=255pt]{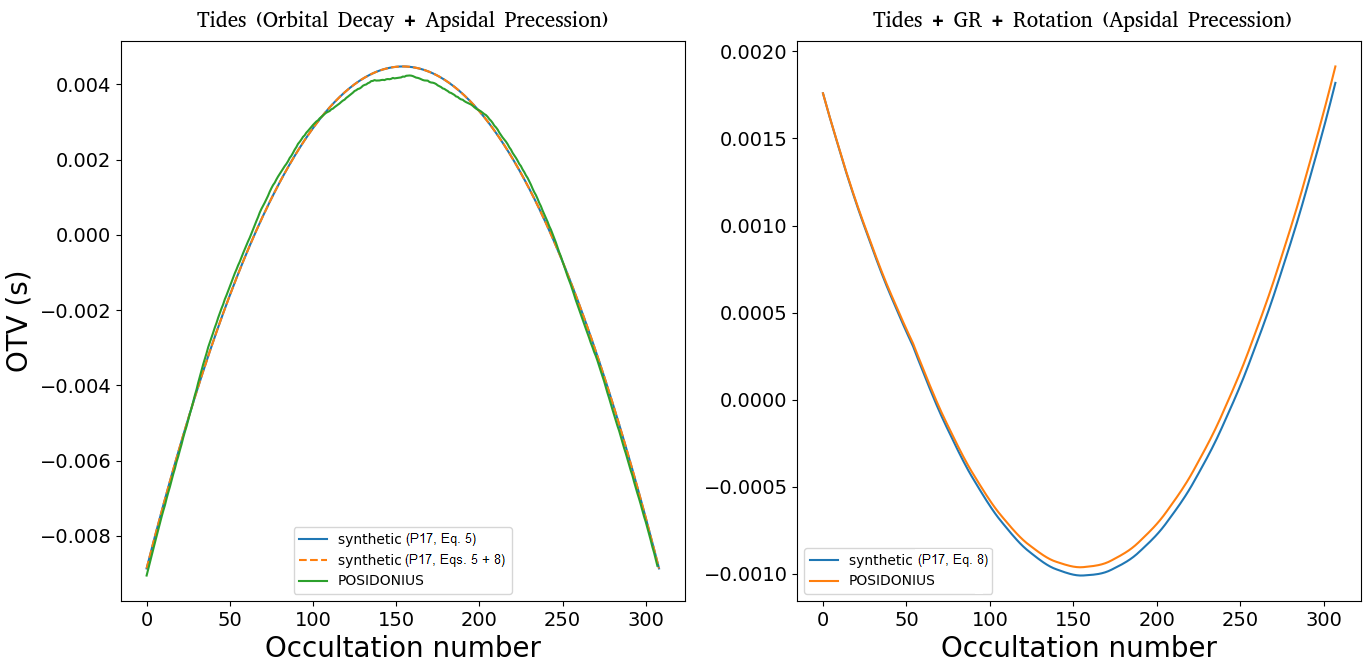}
\caption{Occultation timing variations considering the same simulations of Fig.\,\ref{fig:fig9}, i.e., a K2-265 b with $e=0.15$, a synchronous rotation rate and a viscosity of $\eta = 10^{22} \ \textrm{Pa s}$. The panel on the left shows the contribution of orbital decay and tidally-induced apsidal precession to the OTVs, while the panel on the right shows the contribution of apsidal precession induced by three effects to the OTVs (i.e., for the latter, the effects of GR, tides and rotation were considered). The synthetic OTV curves were generated by considering Eqs. 5 and 8 given in Section 4 of \citet{Patra2017}, corresponding to P17 in the labels of the figure.}
\label{fig:fig10}
\end{figure}

Although GR has been shown to play a non-negligible role on the TTVs only in the case of synchronous rotation, it is possible that GR and tides provide the same effects on TTVs if the mass and radius of the planet under study lead to relatively small tidally-induced TTVs. In such specific cases, the interplay between GR and tides and their influences on the light curves of exoplanets can be studied by analysing occultation curves as well as transit curves. As it was discussed by e.g., \citet{Patra2017} and \citet{Yee2020}, the shape of the timing variation curves of transits and occultations differ when considering the contributions of apsidal precession and orbital decay. Fig.\,\ref{fig:fig10} shows an example of occultation timing variations (OTVs) curves corresponding to a homogeneous K2-265 b on a synchronous rotation rate regime with $e=0.15$ and $\eta = 10^{22} \ \textrm{Pa s}$.

Analysing the contributions coming from orbital decay and apsidal precession to the OTVs shown in Fig.\,\ref{fig:fig10}, we can see that the occultations provide a way to disentangle between these effects when they have comparable contributions to the timing variations. We emphasize that the use of occultation timing analysis to disentangle between apsidal precession and orbital decay-induced timing variations has already been employed in \citet{Yee2020} to confirm the orbital decay of WASP-12 b, and we briefly present the curves in Fig.\,\ref{fig:fig10} to show that tidal interactions and GR can be analysed separately if occultation data are available (i.e., there is no confusion between the two contributions if both transits and occultation curves are available). However, we emphasize that the only case for which apsidal precession can have a non-negligible effect on the timing curves corresponds to a synchronous rotation, in which case the orbital decay rate is much smaller when compared to non-synchronous resonant states. In such specific case, model comparison tools (such as the analysis of the Bayesian Information Criterion, see discussions in \citealt{Patra2017} and \citealt{Yee2020}) have to be employed to choose between models with different numbers of free parameters (e.g., a model considering a fixed orbital decay rate and a model considering an eccentric orbit with fixed values for the apsidal precession rate and argument of periapsis).

\subsection{Broad exploration of parameters space}

The analysis performed in Sec.\,\ref{sec:sec4} regarding the broad exploration of tidally-induced TTVs as a function of both planetary and stellar parameters has allowed for the conclusion that the main components leading to significant changes in TTV amplitudes are the stellar mass, planetary radius and semi-major axis. The planetary mass was shown to present little influence on the amplitudes of the tidally-induced TTVs. Moreover, the results presented in Sec.\,\ref{sec:sec4} showed that the most promising planetary systems for which tidally-induced TTVs would be detectable are the ones for which the orbital period is of the order 2 days or smaller, and in which the planet can have a moderate eccentricity of the order $0.1$ or higher, thus allowing for non-synchronous spin-orbit resonant states to be maintained until the present, thus enhancing the amplitude of tidally-induced TTVs when compared to the synchronous rotation rate case. 

We emphasize that all calculations presented in this work for the tidally-induced TTVs considered that the planet adjusts to hydrostatic equilibrium following a Newtonian creep equation for its shape evolution. We also neglected forced librations of the rotation rate by considering the constant rotation rate approximation of the creep tide theory. \footnote{For planets trapped in spin-orbit resonances, forced librations can have an important role on the tidal heating of the planets (see e.g., discussions in \citealt{Efroimsky2018,Correia2019}).} Considering that the planet has permanent components of the flattenings as a consequence of e.g., reorientation or despinning (see e.g., \citealt{Matsuyama2008,Matsuyama2009}) may lead to even bigger TTVs\footnote{To take into account permanent components of the flattenings, a different implementation of tidal interactions must be used. In the frame of the creep tide, for instance, we would have to solve three additional first-order differential equations dictating the shape evolution of the body instead of supposing the constant rotation rate approximation, which automatically neglects forced librations of the rotation rate (see e.g., discussions in \citealt{Folonier2018,Gomes2019}).}. Detecting these permanent shape-induced TTVs may be essential at characterizing planetary interior structures as well as studying their past orbital and rotational configuration.

\section{Conclusion}

\label{sec:concl}

The most important findings of this work regard the analysis of the tidally-induced TTVs for non-synchronous spin-orbit resonant states. For a given eccentricity value, non-synchronous spin-orbit resonances lead to a faster tidally-induced orbital evolution process when compared to synchronous rotation rate cases. As a consequence, planets in non-synchronous spin-orbit resonances migrate faster and may present larger TTVs than the ones generally predicted by employing classical expressions to calculate the orbital decay rate induced by tides (we refer to classical expressions as the ones generally based on the CTL model, which assumes that the only possible equilibrium rotation state is pseudo-synchronism). Moreover, we have discussed that, when occultation timing data are available in addition to transit timing data, the potential degenerescence of tidal interactions with other effects inducing timing variations (such as general relativity and stellar rotation) can be broken by analysing the occultation timing variations. We also emphasize that in all cases of non-synchronous rotation, the orbital decay-induced TTV is at least 2 orders of magnitude bigger than the apsidal precession-induced TTV.

Another discussion that ensued from this work is the possibility of future detection of tidally-induced TTVs caused by orbital decay for other exoplanetary systems containing a close-in rocky planet with a non-negligible eccentricity. A quick estimation of the tidally-induced TTVs as a function of the semi-major axis in the case of a moderately-eccentric planet (with $0.1 < e < 0.2$) in the 3/2 spin-orbit resonance allows us to conclude that, for planets with $a \leq 0.02 \ \textrm{AU}$, an amplitude of the tidally-induced TTVs of the order $20 - 80 \ \textrm{s}$ may be reached even for small observation timescales (of the order of $2 - 3$ years), provided that the stellar mass is between $0.5$ and $1.0$ Solar masses. Some of the (putative) single-planet systems in which the planet satisfies such semi-major axis criterion include, for instance, LHS-3844 b \citep{Vanderspek2019} and L 168-9 b \citep{Astudillo2020}. Since very close-in planets are believed to have small eccentricities due to tidally-induced orbital circularization processes, the search for the detection of tidally-induced TTVs for Earth-like rocky planets may be more advantageous for planetary systems with at least one more planetary companion. In this case, eccentricity excitation as a consequence of planet-planet gravitational interactions could lead to moderate eccentricities for the inner planet. Consequently, non-synchronous spin-orbit resonances may be maintained until the present. Some of the multi-planet systems which can present this configuration are K2-38 b-c \citep{Sinukoff2016,Toledo2020}, LTT 3780 b-c \citep{Nowak2020,Cloutier2020} and TRAPPIST-1 \citep{Gillon2017}. The new version of the Posidonius code introduced in this work (which has been shown to provide stable and precise results) can thus be a powerful tool for studying the cases of these multiplanetary systems, for which analytical formulations of the transit and occultation timings may not be able to capture all the aspects of planet-planet perturbations.

\begin{acknowledgements}
The authors would like to thank the anonymous referee for helping improving the manuscript. G.O.G would like to thank FAPESP for funding the project under grants 2016/13750-6, 2017/25224-0 and 2019/21201-0. This work has been carried out within
the framework of the NCCR PlanetS supported by the Swiss National Science Foundation. This research has made use of NASA’s Astrophysics Data System.
\end{acknowledgements}

\bibliographystyle{aa} 
\bibliography{Biblio} 

\begin{appendix}
\section{Code verification and performance}

The Posidonius code considers the effects of additional forces other than N-body point-mass interactions by directly implementing the contribution of these effects in the force components acting on each body of a given system. Both a symplectic integration scheme with a fixed time-step (WHFast integrator) and an integrator with a  variable time-step integration scheme (IAS15) are available for use in the code. 

To test the proper functioning of the implementation of the creep tide theory in the Posidonius code, we compared the results coming from it with the ones coming from the use of a secular code giving the spin-orbit evolution of the body. The equations for the secular evolution are

\begin{equation}
\dot{a} = \frac{R^2 n \bar{\epsilon}_{\rho}}{5a} \sum_{k \in \mathbb{Z}} \left[3(2-k) \frac{\gamma (\nu + kn)E_{2,k}^2}{\gamma^2 + (\nu+kn)^2} - \frac{\gamma k^2 n E_{0,k}^2}{\gamma^2 + k^2n^2}\right] ,
\label{eq:dot-a-nsynch-visc}
\end{equation}

\begin{equation}
\dot{e} = -\frac{3GMR^2 \bar{\epsilon}_{\rho}}{10na^5e} \sum_{k \in \mathbb{Z}} \left[P_k^{(1)} \frac{\gamma (\nu + kn)E_{2,k}^2}{\gamma^2 + (\nu+kn)^2} +\frac{P^{(2)}}{3} \frac{\gamma k^2 n E_{0,k}^2}{\gamma^2 + k^2n^2}\right] ,
\label{eq:dot-e-nsynch-visc}
\end{equation}

\begin{equation}
\dot{\Omega} = -\frac{3GM\bar{\epsilon}_{\rho}}{2a^3} \sum_{k \in \mathbb{Z}} \frac{\gamma (\nu + kn)E_{2,k}^2}{\gamma^2 + (\nu+kn)^2},
\label{eq:dot-omega-nsynch-visc}
\end{equation}
where $\nu = 2\Omega - 2n$ is the semidiurnal frequency of the primary, the coefficients $P_k$ are eccentricity-dependent coefficients given by
\begin{equation}
P_k^{(1)} = 2\sqrt{1-e^2} - (2-k) (1-e^2) , 
\end{equation}
\begin{equation}
P^{(2)} = 1-e^2 ,    
\end{equation}
and the coefficients $E_{q,k}$ are the eccentricity-dependent Cayley coefficients. We can see that no short-period components (i.e., components depending on the true or mean anomaly) exist, since they were already averaged out.

\begin{figure}
\centering
\includegraphics[height=230pt,width=250pt]{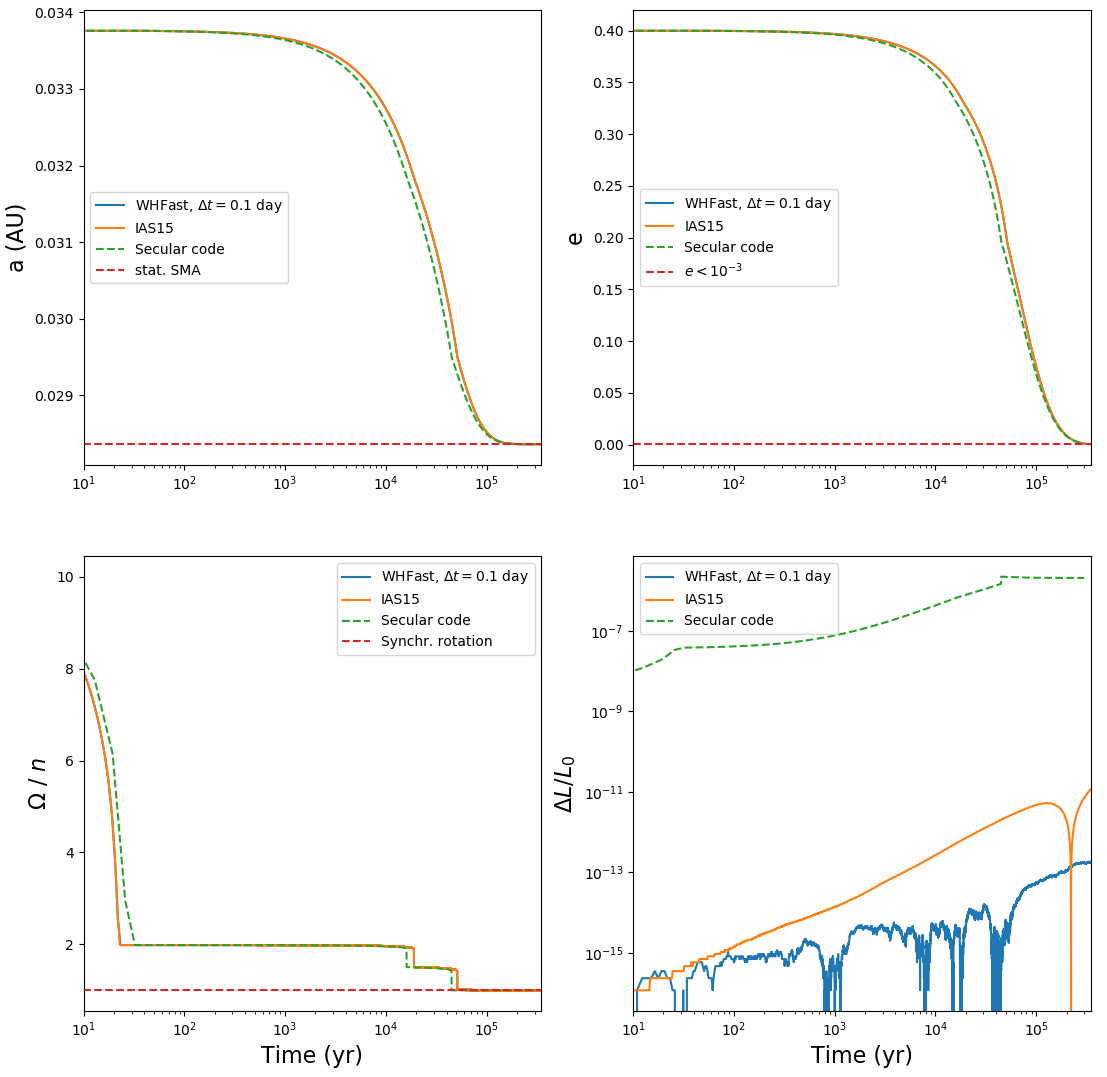}
\caption{Semi-major axis, eccentricity, rotation and angular momentum evolution of four simulations (see the labels in the panels). Solid lines correspond to simulations using the Posidonius code, and the green dashed line corresponds to the simulation using the secular code. The red dashed line shown in the panels of the orbital and rotational evolution of the planet corresponds to the predictions of the final values of the spin-orbit configuration of the planet, based on total angular momentum conservation.}
\label{fig:fig8}
\end{figure}

Figure\,\ref{fig:fig8} shows an example of the results of three simulations considering the spin and orbit evolution of a homogeneous K2-265 b with a viscosity of $\eta = 10^{17} \ \textrm{Pa s}$ (the choice to consider such value for $\eta$ relies on the fact that lower values of $\eta$ lead to bigger amplitudes of the short-period oscillations of $\Omega$, thus representing the best scenario for identifying potential changes introduced by the averaging of the equations of motion, which are performed in the frame of the secular model). 

We can see several characteristics which are classic of tidal evolution scenarios in Fig.\,\ref{fig:fig8}: Firstly, the rotation of the planet is damped to a 2/1 spin-orbit resonant state. This process is followed by orbital shrinking and eccentricity damping until the 2/1 spin-orbit resonance is no longer stable. The rotation then evolves to the 3/2 spin-orbit resonance while the eccentricity and semi-major axis continue to decrease (we emphasize that the evolution of the orbit takes place on a timescale much bigger than the evolution of the rotation). The endpoint of tidal evolution is achieved when both orbital circularization and rotational synchronization take place. We can see that the values of the orbital elements at the endpoint of the tidal evolution are in very good agreement with the analytical estimations coming from the angular momentum conservation of the system, namely $a_{\textrm{stat}} = a_0 (1-e_0^2)$ (see red dashed curve on the top panel on the left in Fig.\,\ref{fig:fig8}). 

From the point of view of angular momentum conservation, we can see that the Posidonius code conserves it with a much better precision when compared to the secular code (see bottom panel on the right in Fig.\,\ref{fig:fig8}). Additionally, we verified that the WHFast code better conserves the angular momentum for a sufficiently long timescale, whereas the IAS15 integrator gives more precise results for short-term evolution scenarios. Lastly, we comment that the speed of integration is much slower for the IAS15 when compared to the WHFast integrator. The ratio of time taken to evolve the orbits with the IAS15 and WHFast integrators is approximately 25, which is an expected difference in integration time given the adaptative time-step scheme of the IAS15 integrator. We comment that the value of the (constant) time-step chosen for the WHFAST was set so that it would be near the mean value of the time-step chosen by the adaptative time-step algorithm of the IAS15 integrator (for the latter case, we verified that the time-step oscillated between 0.08 and 0.13 days for the simulation shown in Fig.\,\ref{fig:fig8}). We also performed some experiments by tuning the time-step of the WHFast integrator and verified that the results obtained by considering a time-step smaller than 0.1 days do not change when compared to the case in which the time-step was chosen to be 0.1 days (i.e., the case corresponding to the blue curve in Fig.\,\ref{fig:fig8}). For a more comprehensive discussion on the precision and error analyses for symplectic integrators, see \citet{Hernandez2017}.

\end{appendix}

\end{document}